\documentclass[11pt,a4paper]{article}
\usepackage{amsmath}
\usepackage{amssymb}
\usepackage{graphicx}
\usepackage{booktabs}
\usepackage{float}
\usepackage{tikz}
\usepackage{pgfplots}
\usepackage{authblk} 
\pgfplotsset{compat=1.18}
\newcommand{\dd}{\mathrm{d}}
\newcommand{\Iinv}{\mathcal{I}}
\newcommand{\LAdS}{L_{AdS_2}}
\begin{document}
	\title{Superluminal propagation in the magnetic NED reconstruction
		of the extremal Dymnikova black hole}
	\author{J. M. Romero\thanks{Corresponding author}}
	\affil{IFIMAR--CONICET, Mar del Plata, Argentina}
	\date{Received: date / Revised version: date}
	\maketitle
	\begin{abstract}
		We reconstruct the magnetic nonlinear electrodynamics (NED) source of the
		Dymnikova geometry on the branch \(\Iinv>0\) and analyze its
		electromagnetic perturbations on the fixed background. The reconstruction
		does not admit a Maxwell limit. The extraordinary characteristic cone
		gives a transverse speed \(v_\perp^2=(3u-5)/2\), with \(u=r^3/P^3\),
		which becomes superluminal for \(r>r_{\rm lum}=(7/3)^{1/3}P\). The
		principal system is pointwise hyperbolic in the finite exterior
		\(r\ge r_e\), with positive local quadratic energy, but not uniformly
		hyperbolic up to spatial infinity. For axial perturbations, the radial effective potential takes the
		closed form \(V_{\rm em}=f\ell(\ell+1)\mathcal{A}/r^2
		+(\sqrt{\mathcal{L}_{\Iinv}})_{,r_*r_*}/\sqrt{\mathcal{L}_{\Iinv}}\),
		vanishes as \((r-r_e)^2\) at the extremal horizon, and grows as
		\(9r^4/(4P^6)\) at infinity. Its \(\ell\)-independent part is negative
		in a bounded region, but the potential is numerically positive for all
		\(\ell\ge1\) on a fine grid up to \(r=50P\), with a uniform margin. The Dymnikova reconstruction
		does not satisfy the hypotheses of the Tomizawa--Suzuki theorem, since it
		lacks a Maxwell limit, yet it still produces superluminal extraordinary
		photons. The reconstructed NED branch is therefore mathematically
		consistent as a source of the Dymnikova background, but not causally
		acceptable with respect to the physical metric in the asymptotic region.
		The global positivity of the effective potential and the quasinormal
		spectrum are left as open problems.
		
		\vspace{0.5em}
	\end{abstract}
	\maketitle
	\section{Introduction}
	\label{sec:intro}
	
	The original Dymnikova regular black hole \cite{Dymnikova1992}, which we refer to as Dymnikova-I to distinguish it from later metrics introduced by the same author \cite{Dymnikova2004}, has been studied in
	several contexts, including generalized uncertainty principle corrections
	\cite{GUP}, dark matter and quintessence \cite{Quintessence}, and
	quasinormal modes \cite{QNM}. The same geometry has also been derived
	in an iterative renormalization-group approach \cite{Platania2019}. Its magnetic NED reconstruction was
	considered by Alencar and Borralho as an auxiliary step in their analysis
	of unimodular gravity \cite{AlencarGRG2026}. Magnetic monopoles in NED
	have also been considered in Ref.~\cite{Bronnikov2001}.
	
	This work is deliberately restricted in scope. The extremality condition
	\(e^{u_*}=1+3u_*\), the value \(u_*\simeq1.9038\), and the extremal mass
	\(M_*\simeq0.7282\,P/G\) are algebraic consequences of the extremality
	conditions; we derive them explicitly here but do not claim them as new
	results. The magnetic NED reconstruction was constructed in Sec.~II of
	Ref.~\cite{AlencarGRG2026}, and the violation of the standard energy
	conditions in the pure Dymnikova case is already noted in
	Ref.~\cite{GUP}. Our contribution is the explicit computation of the
	near-horizon \(AdS_2\) radius, the linearized electromagnetic stability
	analysis of the magnetic branch, a closed form of the radial effective
	potential for axial perturbations, and a quantitative example of superluminal
	propagation in a NED model without a Maxwell limit. The latter is
	discussed in relation to the Tomizawa--Suzuki theorem
	\cite{TomizawaSuzuki}.
	
	The paper is organized as follows. In Sec.~\ref{sec:geometry} we review
	the extremal geometry and compute \(L_{AdS_2}\). In Sec.~\ref{sec:source}
	we give the Einstein tensor and the energy-condition classification,
	including the comparison with previous results and with the
	Tomizawa--Suzuki theorem. In Sec.~\ref{sec:NED} we present the magnetic
	NED reconstruction with explicit conventions. In
	Sec.~\ref{sec:stability} we analyze the linearized electromagnetic
	perturbations on a fixed background, derive the extraordinary cone, and
	derive the radial effective potential in closed form. In
	Sec.~\ref{sec:conclusions} we summarize and state the open problems.
	Appendix~\ref{app:comparison} gives a term-by-term comparison with the
	conventions of Ref.~\cite{AlencarGRG2026}.
	
	\section{Extremal geometry and the \(AdS_2\) radius}
	\label{sec:geometry}
	
	The Dymnikova-I metric \cite{Dymnikova1992} with \(r_0=P\) is
	\begin{equation}
		ds^2=-f(r)\dd t^2+\frac{\dd r^2}{f(r)}+r^2\dd\Omega_2^2,
		\qquad
		f(r)=1-\frac{2GM}{r}\left(1-e^{-r^3/P^3}\right).
		\label{eq:metric}
	\end{equation}
	We use \(m\equiv GM\) and \(u\equiv r^3/P^3\). The extremality conditions
	\(f(r_e)=f'(r_e)=0\) give
	\begin{equation}
		e^{u_*}=1+3u_*,
		\qquad
		u_*\simeq1.9038137,
		\qquad
		r_e\simeq1.2393905\,P.
		\label{eq:extremality}
	\end{equation}
	These values follow algebraically from Eq.~\eqref{eq:metric}; we quote
	them as input for the rest of the analysis.
	
	Near \(r_e\), \(f(r)\approx\frac12 f''(r_e)(r-r_e)^2\), and the metric
	approaches \(AdS_2\times S^2\) with
	\begin{equation}
		\LAdS^2=\frac{2}{f''(r_e)}.
		\label{eq:LAdS_def}
	\end{equation}
	A direct computation gives
	\begin{equation}
		f''(r_e)=\frac{3u_*-2}{r_e^2},
		\qquad
		\boxed{
			\LAdS=r_e\sqrt{\frac{2}{3u_*-2}}\simeq0.7341\,r_e.
		}
		\label{eq:LAdS}
	\end{equation}
	This is a direct algebraic consequence of Eq.~\eqref{eq:extremality}. We
	present it as an explicit geometric result, not as an independent
	conceptual contribution.
	
	\section{Einstein tensor and energy conditions}
	\label{sec:source}
	
	The non-zero diagonal mixed components of the Einstein tensor are
	\begin{equation}
		G^t{}_t=G^r{}_r=-\frac{6m}{P^3}e^{-u},
		\qquad
		G^\theta{}_\theta=\left(-\frac{6m}{P^3}+\frac{9mr^3}{P^6}\right)e^{-u}.
		\label{eq:G_components}
	\end{equation}
	The physical density and pressures, defined through
	\(G_{\mu\nu}=8\pi G\,T_{\mu\nu}^{\rm phys}\), are
	\begin{equation}
		\rho_{\rm phys}=\frac{6m}{8\pi GP^3}e^{-u},
		\qquad
		p_{r,{\rm phys}}=-\rho_{\rm phys},
		\qquad
		p_{t,{\rm phys}}=\rho_{\rm phys}\left(-1+\frac{3u}{2}\right).
		\label{eq:rho_pressures}
	\end{equation}
	
	The NEC combinations are \(\rho+p_r=0\) (radial saturation) and
	\(\rho+p_t=(3u/2)\rho\ge0\) (tangential, satisfied). The WEC holds
	everywhere. The SEC combination is
	\begin{equation}
		\rho_{\rm phys}+p_{r,{\rm phys}}+2p_{t,{\rm phys}}=(3u-2)\rho_{\rm phys},
		\label{eq:SEC}
	\end{equation}
	violated for \(u<2/3\), saturated at \(u=2/3\), satisfied for \(u>2/3\).
	The DEC requires \(\rho\ge|p_t|\), equivalent to
	\begin{equation}
		0\le u\le\frac43.
		\label{eq:DEC}
	\end{equation}
	At extremality, \(u_*\simeq1.9038>4/3>2/3\), so the SEC is satisfied at
	the horizon and outside, while the DEC is violated there. The transition
	radii are \(r_{\rm SEC}=(2/3)^{1/3}P\) and \(r_{\rm DEC}=(4/3)^{1/3}P\),
	with \(r_{\rm DEC}<r_e\).
	
	\paragraph{Comparison with previous results.}
	Dymnikova has introduced several distinct regular black-hole metrics. The
	geometry \eqref{eq:metric} considered here is the original de Sitter-core
	model of Ref.~\cite{Dymnikova1992} (Dymnikova-I). For this geometry the
	DEC is violated for \(u>4/3\), since \(\rho-p_t=\rho(2-3u/2)<0\) there,
	and, at extremality, at the horizon and throughout the exterior. This is
	consistent with the violation of the standard energy conditions noted in
	Ref.~\cite{GUP}. Ref.~\cite{Maeda} shows that the Dymnikova black hole
	satisfies the DEC everywhere; that analysis refers to a different metric,
	the nonlinear-electrodynamics solution of Ref.~\cite{Dymnikova2004}
	(Dymnikova-II), with mass function
	\(M(r)=\frac{2m}{\pi}\left[\arctan(r/l)-lr/(r^2+l^2)\right]\), and not to
	the Dymnikova-I geometry. The DEC used in Ref.~\cite{Maeda} is equivalent
	to the standard Hawking--Ellis condition \(\rho\ge|p_i|\). There is
	therefore no contradiction between the two results. In terms of the
	criteria proposed in Ref.~\cite{Maeda}, the Dymnikova-I geometry fails the
	requirements that the standard energy conditions hold in the asymptotically
	flat region and on the horizon of large black holes, since the DEC is
	violated for all \(r>r_{\rm DEC}\); this is consistent with the choice of
	the Dymnikova-II metric in that work. The optical properties of the
	Dymnikova-II solution in its effective geometry have been studied in
	Ref.~\cite{dePaulaDymnikova}.
	
	\paragraph{Relation to the Tomizawa--Suzuki theorem.}
	Tomizawa and Suzuki proved that, in static spherically symmetric
	spacetimes and for any NED admitting the Maxwell limit, timelike
	propagation of photons with nonzero angular momentum implies the DEC; the
	converse holds at least in the weak-field limit \cite{TomizawaSuzuki}. Their
	theorem requires the Maxwell limit as a hypothesis. The Dymnikova
	reconstruction considered here does not admit a Maxwell limit, since both
	\(\mathcal{L}\) and \(\mathcal{L}_{\Iinv}\) vanish as
	\(\Iinv\to0^+\). The theorem therefore does not apply, and no
	contradiction with Ref.~\cite{TomizawaSuzuki} arises.
	
	In NED, the photon 4-momentum \(k_\mu\) and the ray 4-velocity
	\(u^\mu=g_{\rm eff}^{\mu\nu}k_\nu\) are not parallel, and their norms with
	respect to the physical metric can have opposite signs; causality must be
	assessed with \(u^\mu\) \cite{dePaulaTachyons}. For the extraordinary mode
	of Sec.~\ref{sec:stability} propagating transversally,
	\(g_{\mu\nu}u^\mu u^\nu=\mathcal{A}(\mathcal{A}-1)k_\perp^2\) and
	\(g^{\mu\nu}k_\mu k_\nu=(1-\mathcal{A})k_\perp^2\). For
	\(r>r_{\rm lum}\) the rays are spacelike while the 4-momentum is
	timelike: in the terminology of Ref.~\cite{dePaulaTachyons}, the
	superluminal photons are bradyonic (``bad bradyons''). As in the general
	analysis of that reference, the sign of \(\mathcal{L}_{\Iinv\Iinv}\)
	controls causality. Here
	\(\bar\Iinv\mathcal{L}_{\Iinv\Iinv}/\mathcal{L}_{\Iinv}=(3u-7)/4\), so the
	Lagrangian is concave for \(u<7/3\) and convex for \(u>7/3\), and the
	change of sign coincides with \(r_{\rm lum}\). Likewise, the sufficient
	DEC conditions \(\mathcal{L}\ge0\), \(\mathcal{L}_{\Iinv}\ge0\) and
	\(\mathcal{L}-\Iinv\mathcal{L}_{\Iinv}\ge0\) \cite{dePaulaTachyons} give
	\(\mathcal{L}-\Iinv\mathcal{L}_{\Iinv}=\mathcal{L}(1-3u/4)\ge0\), which
	reproduces Eq.~\eqref{eq:DEC}. The Dymnikova-I reconstruction therefore
	joins the Bardeen and Hayward models in the class with a mixed sign of
	\(\mathcal{L}_{\Iinv\Iinv}\), which violates the DEC and is acausal
	\cite{dePaulaTachyons}.
	
	With causality assessed through \(u^\mu\), the two transitions are
	distinct. For \(r_{\rm DEC}<r<r_{\rm lum}\), a range that contains the
	near-horizon exterior \(r_e\le r<r_{\rm lum}\), the DEC is violated but
	propagation is causal. This is an explicit realization of the remark of
	Tomizawa and Suzuki that, without the Maxwell limit, a violation of the
	DEC does not necessarily imply spacelike photon propagation
	\cite{TomizawaSuzuki}. For \(r>r_{\rm lum}\), the DEC is violated and the
	rays are superluminal. This provides a quantitative example of
	superluminal propagation in a NED model outside the Maxwell-limit class.
	It is also consistent with the general result that causality implies the
	DEC in nonlinear electrodynamics \cite{RussoTownsend}: the reconstructed
	theory violates the DEC and is acausal in its weak-field region.
	
	\section{Magnetic NED reconstruction}
	\label{sec:NED}
	
	For the magnetic monopole,
	\begin{equation}
		F=P\sin\theta\,\dd\theta\wedge\dd\phi,
		\qquad
		\Iinv\equiv F_{\mu\nu}F^{\mu\nu}=\frac{2P^2}{r^4}>0.
		\label{eq:monopole}
	\end{equation}
	We adopt the convention \(\Iinv=F_{\mu\nu}F^{\mu\nu}\) without the
	\(1/4\) factor, the metric signature \((-,+,+,+)\), and the action
	\begin{equation}
		S=\frac{1}{16\pi G}\int\dd^4x\sqrt{-g}\left[R-\mathcal{L}(\Iinv)\right].
		\label{eq:action}
	\end{equation}
	The resulting Einstein equations are
	\begin{equation}
		G_{\mu\nu}=2\mathcal{L}_{\Iinv}F_{\mu\lambda}F_\nu{}^\lambda
		-\frac12 g_{\mu\nu}\mathcal{L}.
		\label{eq:einstein_NED}
	\end{equation}
	Matching the temporal component \(G^t{}_t=-\mathcal{L}/2\) to
	Eq.~\eqref{eq:G_components} gives
	\begin{equation}
		\mathcal{L}(\Iinv)=\frac{12m}{P^3}
		\exp\left[-\left(\frac{2}{P^2\Iinv}\right)^{3/4}\right],
		\qquad
		\mathcal{L}_{\Iinv}=\frac{9mr^7}{2P^8}e^{-u}.
		\label{eq:Lagrangian}
	\end{equation}
	The reconstruction is restricted to \(\Iinv>0\), the domain sampled by
	the monopole background. We do not assume an extension to
	\(\Iinv\le0\); such an extension would be additional information not
	determined by the magnetic solution.
	
	The reconstruction coincides with the magnetic branch of
	Ref.~\cite{AlencarGRG2026}, with the difference of conventions discussed
	in Appendix~\ref{app:comparison}. It does not reduce to Maxwell theory at
	weak field, since \(\mathcal{L}\to0\) and \(\mathcal{L}_{\Iinv}\to0\) as
	\(\Iinv\to0^+\), and it depends explicitly on both \(m\) and \(P\), so it
	is a solution-dependent reconstruction rather than a universal NED
	theory.
	
	\section{Electromagnetic perturbations on a fixed Dymnikova background}
	\label{sec:stability}
	
	\subsection{Linearized equations and principal system}
	
	Writing \(F_{\mu\nu}=\bar F_{\mu\nu}+f_{\mu\nu}\), with \(f_{\mu\nu}\) of first order, and
	\(\Iinv=\bar{\Iinv}+\delta\Iinv\) with
	\(\delta\Iinv=2\bar F^{\alpha\beta}f_{\alpha\beta}\), the linearized NED
	equation on the fixed background is
	\begin{equation}
		\bar\nabla_\mu\left[
		\mathcal{L}_{\Iinv}f^{\mu\nu}
		+2\mathcal{L}_{\Iinv\Iinv}
		\left(\bar F^{\alpha\beta}f_{\alpha\beta}\right)\bar F^{\mu\nu}
		\right]=0,
		\label{eq:lin_NED}
	\end{equation}
	together with the linearized Bianchi identity. In a local orthonormal
	frame with the \(x\)-axis along \(\bar{\mathbf B}\), the constitutive
	relations are
	\begin{equation}
		\mathbf d=\mathcal{L}_{\Iinv}\mathbf e,
		\qquad
		\mathbf h=\mathcal{L}_{\Iinv}\mathbf b
		+4\mathcal{L}_{\Iinv\Iinv}(\bar{\mathbf B}\cdot\mathbf b)\bar{\mathbf B}.
		\label{eq:constitutive}
	\end{equation}
	Separating parallel and perpendicular components,
	\begin{equation}
		h_x=\mathcal{C}b_x,
		\qquad
		h_y=h_z=\mathcal{L}_{\Iinv}b_{y,z},
		\qquad
		\mathcal{C}\equiv\mathcal{L}_{\Iinv}+2\bar{\Iinv}\mathcal{L}_{\Iinv\Iinv}.
		\label{eq:C_def}
	\end{equation}
	
	\subsection{Characteristic cones}
	
	In the eikonal approximation, the condition for nontrivial solutions
	separates into two polarization sectors. The first (ordinary)
	propagates on the metric light cone \(\bar g^{\mu\nu}k_\mu k_\nu=0\). The
	second (extraordinary) propagates on
	\begin{equation}
		g_{\rm eff}^{\mu\nu}k_\mu k_\nu=0,
		\qquad
		g_{\rm eff}^{\mu\nu}=\bar g^{\mu\nu}
		+\frac{4\mathcal{L}_{\Iinv\Iinv}}{\mathcal{L}_{\Iinv}}
		\bar F^{\mu\alpha}\bar F^\nu{}_\alpha.
		\label{eq:eff_metric}
	\end{equation}
	In the local orthonormal frame,
	\begin{equation}
		g_{\rm eff}^{\hat a\hat b}={\rm diag}(-1,1,\mathcal{A},\mathcal{A}),
		\qquad
		\mathcal{A}=\frac{\mathcal{C}}{\mathcal{L}_{\Iinv}}
		=1+\frac{2\bar{\Iinv}\mathcal{L}_{\Iinv\Iinv}}{\mathcal{L}_{\Iinv}}.
		\label{eq:eff_metric_diag}
	\end{equation}
	For our Lagrangian,
	\begin{equation}
		\boxed{\mathcal{A}(u)=\frac{3u-5}{2}.}
		\label{eq:A_u}
	\end{equation}
	The extraordinary dispersion relation is
	\(\omega^2=k_r^2+\mathcal{A}k_\perp^2\). Radial propagation is luminal;
	transverse propagation has speed \(v_\perp=\sqrt{\mathcal{A}}\). The
	splitting into an ordinary and an extraordinary sector is the
	single-invariant form of vacuum birefringence; for Lagrangians depending
	on both electromagnetic invariants, both polarizations are modified
	\cite{dePaulaBirefringence}.
	
	\subsection{Pointwise hyperbolicity and asymptotic degeneracy}
	
	The conditions \(\mathcal{L}_{\Iinv}>0\) and \(\mathcal{A}>0\) hold for
	all finite \(r\ge r_e\), since \(u\ge u_*\simeq1.9038>5/3\). The principal
	system is therefore pointwise hyperbolic in the finite exterior, with
	positive local quadratic energy
	\begin{equation}
		\mathcal{H}^{(2)}\sim\frac12\left[
		\mathcal{L}_{\Iinv}|\mathbf e|^2
		+\mathcal{L}_{\Iinv}|\mathbf b_\perp|^2
		+\mathcal{C}b_\parallel^2
		\right]>0.
		\label{eq:quadratic_energy}
	\end{equation}
	However, \(\mathcal{A}\to\infty\) and \(\mathcal{L}_{\Iinv}\to0\) as
	\(r\to\infty\), and \(\mathcal{C}=\mathcal{L}_{\Iinv}\mathcal{A}\to0\).
	The system is hyperbolic at each finite point but not uniformly hyperbolic
	on the non-compact exterior. This asymptotic degeneracy is characteristic
	of NED reconstructions without a Maxwell limit \cite{Novello2000}.
	
	\subsection{Superluminal propagation}
	
	The transverse speed crosses unity at
	\begin{equation}
		\mathcal{A}=1
		\quad\Longleftrightarrow\quad
		u=\frac73,
		\qquad
		r_{\rm lum}=\left(\frac73\right)^{1/3}P\simeq1.326\,P.
		\label{eq:r_lum}
	\end{equation}
	Since \(r_e\simeq1.239\,P<r_{\rm lum}\), the transition occurs outside
	the horizon. For \(r_e\le r<r_{\rm lum}\), \(v_\perp<1\); for
	\(r>r_{\rm lum}\), \(v_\perp>1\). At the horizon,
	\(v_\perp(r_e)\simeq0.596\); at infinity, \(v_\perp\to\infty\). Since
	\(\omega^2=\mathcal{A}k_\perp^2\) for transverse propagation, \(v_\perp\)
	is both the phase velocity and the ray velocity \(u^\perp/u^t\) of the
	extraordinary mode, so it measures the physical propagation speed
	\cite{dePaulaTachyons}.
	
	Representative values of the transverse speed in the exterior are given in
	Table~\ref{tab:v_perp}.
	
	\begin{table}[t]
		\centering
		\caption{Transverse speed \(v_\perp=\sqrt{(3u-5)/2}\) of the
			extraordinary electromagnetic mode at representative radii, in units
			\(P=1\). The horizon is at \(r_e\simeq1.239\,P\), the luminal
			transition at \(r_{\rm lum}\simeq1.326\,P\).}
		\label{tab:v_perp}
		\begin{tabular}{cccc}
			\toprule
			\(r/P\) & \(u=r^3/P^3\) & \(\mathcal{A}(u)\) & \(v_\perp\) \\
			\midrule
			\(1.2394\) (horizon) & \(1.9038\) & \(0.3557\) & \(0.596\) \\
			\(1.3260\) (\(r_{\rm lum}\)) & \(2.3333\) & \(1.0000\) & \(1.000\) \\
			\(1.5000\) & \(3.3750\) & \(2.5625\) & \(1.601\) \\
			\(2.0000\) & \(8.0000\) & \(9.5000\) & \(3.082\) \\
			\(3.0000\) & \(27.0000\) & \(38.0000\) & \(6.164\) \\
			\(4.0000\) & \(64.0000\) & \(93.5000\) & \(9.670\) \\
			\bottomrule
		\end{tabular}
	\end{table}
	
	The main result of this work is that the extraordinary electromagnetic
	cone of the reconstructed NED branch extends outside the physical light
	cone for \(r>r_{\rm lum}\). The model is locally hyperbolic and has
	positive quadratic energy, but it is not causally acceptable with respect
	to the physical metric in the asymptotic region. Hyperbolicity and
	causality are distinct properties, and the reconstructed branch
	possesses the first but not the second.
	
	\subsection{Radial equation and effective potential}
	
	Only the axial (odd-parity) sector has a nonvanishing perturbation
	\(f_{\theta\phi}\), which is the only component that couples to
	\(\mathcal{L}_{\Iinv\Iinv}\) through \(\delta\Iinv\). This sector therefore
	carries the extraordinary polarization, and we restrict the analysis to it.
	With a real normalized spherical harmonic \(Y_{\ell m}\), we write
	\begin{equation}
		\delta A_\theta=-\frac{a_{\rm em}}{\sin\theta}\,\partial_\phi Y,
		\qquad
		\delta A_\phi=a_{\rm em}\sin\theta\,\partial_\theta Y,
		\label{eq:axial_ansatz}
	\end{equation}
	with \(\delta A_t=\delta A_r=0\) and \(a_{\rm em}=a_{\rm em}(t,r)\), so that
	\(f_{\theta\phi}=-\ell(\ell+1)a_{\rm em}\sin\theta\,Y\). The variable
	\(a_{\rm em}\) is gauge invariant. Expanding the action
	\eqref{eq:action} to second order in \(f_{\mu\nu}\) on the fixed background
	and integrating over the sphere gives
	\begin{equation}
		S^{(2)}=\frac{\ell(\ell+1)}{8\pi G}\int\dd t\,\dd r
		\Bigl[\frac{\mathcal{L}_{\Iinv}}{f}a_{{\rm em},t}^2
		-f\mathcal{L}_{\Iinv}a_{{\rm em},r}^2
		-\frac{\ell(\ell+1)}{r^2}\,\mathcal{C}\,a_{\rm em}^2\Bigr],
		\label{eq:S2}
	\end{equation}
	whose kinetic term is positive for \(\mathcal{L}_{\Iinv}>0\). The radial
	equation is
	\begin{equation}
		a_{{\rm em},tt}
		-\frac{f}{\mathcal{L}_{\Iinv}}
		\left(f\mathcal{L}_{\Iinv}\,a_{{\rm em},r}\right)_{,r}
		+f\frac{\ell(\ell+1)}{r^2}\mathcal{A}\,a_{\rm em}=0.
		\label{eq:radial_master}
	\end{equation}
	With the tortoise coordinate \(dr_*/dr=1/f\) and
	\(\Phi_{\rm em}=\sqrt{\mathcal{L}_{\Iinv}}\,a_{\rm em}\), this becomes
	\(\Phi_{{\rm em},tt}-\Phi_{{\rm em},r_*r_*}+V_{\rm em}\Phi_{\rm em}=0\),
	with
	\begin{equation}
		V_{\rm em}=f\frac{\ell(\ell+1)}{r^2}\mathcal{A}
		+\frac{(\sqrt{\mathcal{L}_{\Iinv}})_{,r_*r_*}}{\sqrt{\mathcal{L}_{\Iinv}}},
		\label{eq:V_em}
	\end{equation}
	which has the structure found for axial perturbations of NED black holes in
	Ref.~\cite{Toshmatov2018}. In terms of \(s\equiv\ln\mathcal{L}_{\Iinv}\)
	and derivatives with respect to \(r\),
	\begin{equation}
		\boxed{
			V_{\rm em}
			=
			f\frac{\ell(\ell+1)}{r^2}\mathcal{A}
			+\frac12ff_{,r}s_{,r}
			+f^2\Bigl(\frac12s_{,rr}+\frac14s_{,r}^2\Bigr).
		}
		\label{eq:V_em_closed}
	\end{equation}
	In the Maxwell case, \(\mathcal{L}_{\Iinv}=1\), this reduces to the
	standard axial potential \(f\ell(\ell+1)/r^2\). For the Dymnikova
	reconstruction, \(s_{,r}=7/r-3r^2/P^3\) and \(s_{,rr}=-7/r^2-6r/P^3\) are
	polynomial, so Eq.~\eqref{eq:V_em_closed} contains no exponentially growing
	or decaying factors. In units \(P=1\), with \(x=r/P\), \(\mu=m/P\) and a
	prime denoting \(d/dx\), the potential splits into an angular part and an
	\(\ell\)-independent part,
	\begin{align}
		V_\ell(x)&=W_\ell(x)+V_0(x),
		\qquad
		W_\ell=f\,\frac{\ell(\ell+1)(3x^3-5)}{2x^2},
		\nonumber\\
		V_0&=\frac12ff'\Bigl(\frac7x-3x^2\Bigr)
		+f^2\Bigl(\frac{9x^4}{4}-\frac{27x}{2}+\frac{35}{4x^2}\Bigr).
		\label{eq:V_decomposition}
	\end{align}
	Since \(f>0\) and \(\mathcal{A}>0\) in the exterior, \(W_\ell>0\).
	
	\paragraph{Horizon expansion.}
	Near the extremal horizon, set \(\epsilon=x-x_e\). The function
	\(D(x)=x+2\mu(e^{-x^3}-1)\), with \(f=D/x\), has a double zero at \(x_e\),
	with \(D''(x_e)=(3u_*-2)/x_e\), so that \(f=c\,\epsilon^2+O(\epsilon^3)\)
	with \(c=(3u_*-2)/(2x_e^2)\). Using \(s_{,x}(x_e)=(7-3u_*)/x_e\),
	\begin{align}
		V_0&=\frac{(3u_*-2)^2(7-3u_*)}{4x_e^5}\,\epsilon^3+O(\epsilon^4),
		\nonumber\\
		W_\ell&=\ell(\ell+1)\frac{(3u_*-2)(3u_*-5)}{4x_e^4}\,\epsilon^2
		+O(\epsilon^3).
		\label{eq:V0_horizon}
	\end{align}
	Numerically, \(V_0\simeq1.517\,\epsilon^3\) and
	\(W_\ell\simeq0.280\,\ell(\ell+1)\,\epsilon^2\). Since
	\(5/3<u_*<7/3\), both coefficients are positive, and for \(\ell\ge1\) the
	potential vanishes as \(\epsilon^2\) at the horizon, dominated by its angular
	part. This expansion should be used in any numerical evaluation that
	approaches the horizon.
	
	\subsection{Asymptotic behavior}
	
	For \(x\to\infty\), \(f\to1\) and \(f'\simeq2\mu/x^2\) up to exponentially
	small terms. The angular part grows as
	\(W_\ell\simeq\frac32\ell(\ell+1)\,x\), while
	\begin{equation}
		\boxed{
			V_0(x)=\frac94x^4-9\mu x^3+9\mu^2x^2-\frac{27}{2}x+51\mu+O(1/x),
		}
		\label{eq:V0_asymptotic}
	\end{equation}
	so the \(\ell\)-independent part dominates and
	\(V_{\rm em}\sim9r^4/(4P^6)\). The ratio
	\begin{equation}
		R(x)\equiv\frac{4V_0(x)}{9x^4}=1-\frac{4\mu}{x}+\frac{4\mu^2}{x^2}
		+O(x^{-3})
		\label{eq:R_asymptotic}
	\end{equation}
	approaches unity from below. The values of \(R(x)\) at representative
	radii are collected in Table~\ref{tab:R_asymptotic}: the asymptotic
	regime \(R\to1\) is reached only for \(x\gtrsim50\), with an error of
	order \(4\mu/x\). At the horizon, \(x\simeq1.24\), the asymptotic formula
	is not applicable, and the expansion \eqref{eq:V0_horizon} must be used
	instead.
	
	\begin{table}[t]
		\centering
		\caption{Ratio \(R(x)=4V_0(x)/(9x^4)\) of the \(\ell\)-independent part
			of the potential to its leading asymptotic term, at representative
			radii. The asymptotic regime \(R\to1\) is reached slowly, with
			\(R(50)\simeq0.943\).}
		\label{tab:R_asymptotic}
		\begin{tabular}{cc}
			\toprule
			\(r/P\) & \(R(x)\) \\
			\midrule
			\(5.0\)  & \(0.4772\) \\
			\(6.0\)  & \(0.5570\) \\
			\(7.0\)  & \(0.6159\) \\
			\(8.0\)  & \(0.6610\) \\
			\(10.0\) & \(0.7255\) \\
			\(15.0\) & \(0.8138\) \\
			\(20.0\) & \(0.8590\) \\
			\(30.0\) & \(0.9051\) \\
			\(40.0\) & \(0.9284\) \\
			\(50.0\) & \(0.9425\) \\
			\bottomrule
		\end{tabular}
	\end{table}
	
	\subsection{Numerical evaluation and positivity}
	
	For electromagnetic perturbations only modes with \(\ell\ge1\) exist: for
	\(\ell=0\) the axial ansatz \eqref{eq:axial_ansatz} vanishes identically, so
	\(V_0\) alone is not the potential of any physical mode. Since \(W_\ell>0\)
	increases with \(\ell\), \(V_\ell\ge V_1\) for all \(\ell\ge1\), and the
	positivity of \(V_1\) is sufficient. The \(\ell\)-independent part is not
	positive by itself: \(V_0<0\) for \(1.2969<x<1.8983\), with a minimum
	\(V_0\simeq-0.185\) at \(x\simeq1.741\).
	
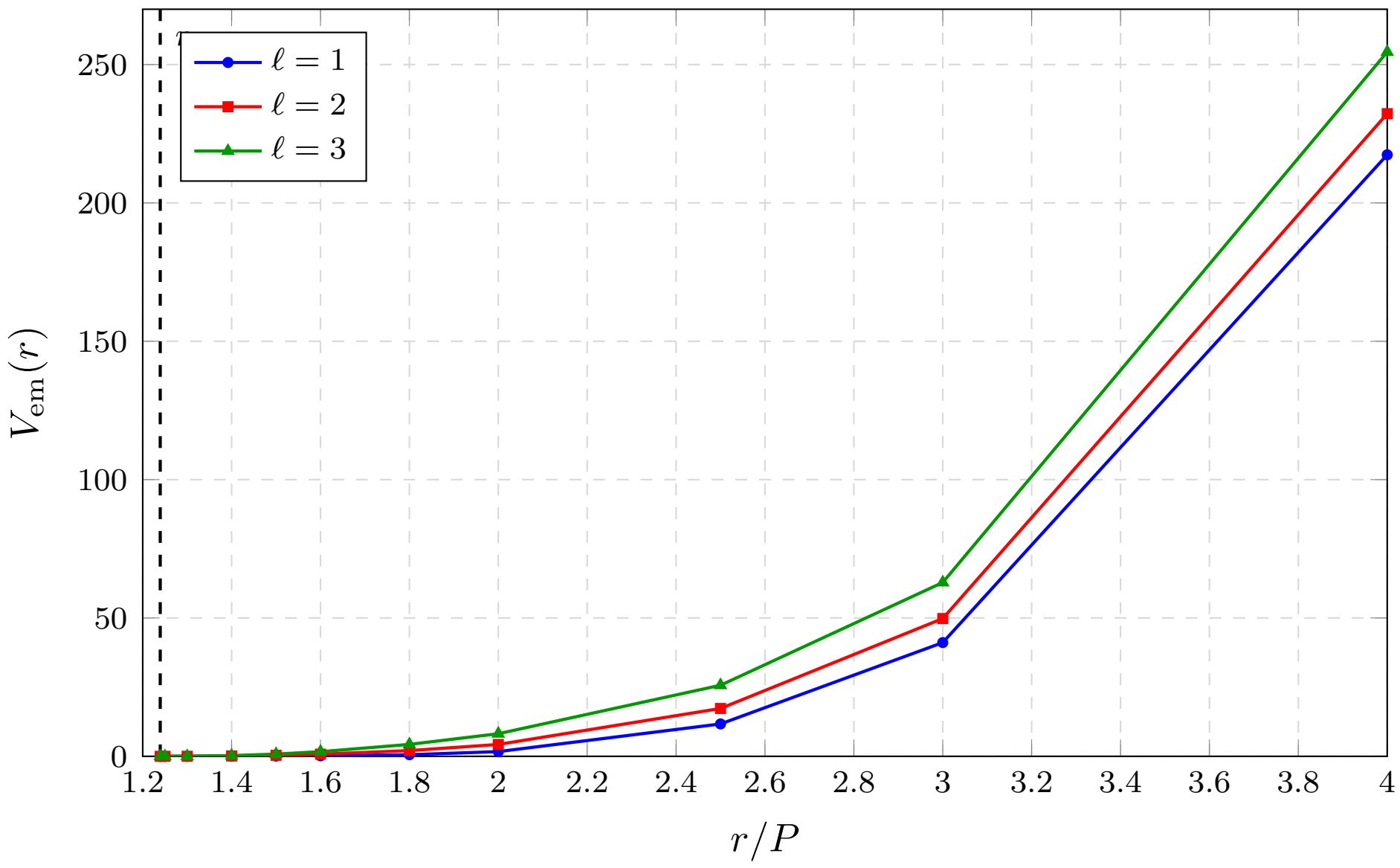
\begin{figure}[t]
		\centering
		\begin{tikzpicture}
			\begin{axis}[
				width=\columnwidth,
				height=0.65\columnwidth,
				xlabel={\(r/P\)},
				ylabel={\(V_{\rm em}(r)\)},
				xmin=1.2, xmax=4.0,
				ymin=0, ymax=270,
				grid=major,
				grid style={dashed, gray!30},
				legend pos=north west,
				legend style={font=\scriptsize},
				tick label style={font=\scriptsize},
				label style={font=\small},
				]
				\addplot[blue, thick, mark=*, mark size=1pt] coordinates {
					(1.2394, 0) (1.2500, 7.55757e-05) (1.3000, 0.0039422) (1.4000, 0.0362594) (1.5000, 0.0965926) (1.6000, 0.18401) (1.8000, 0.581296) (2.0000, 1.70278) (2.5000, 11.6996) (3.0000, 41.1034) (4.0000, 217.397)
				};
				\addlegendentry{\(\ell=1\)}
				\addplot[red, thick, mark=square*, mark size=1pt] coordinates {
					(1.2394, 0) (1.2500, 0.000223827) (1.3000, 0.0118586) (1.4000, 0.124062) (1.5000, 0.380383) (1.6000, 0.781289) (1.8000, 2.07207) (2.0000, 4.28724) (2.5000, 17.2933) (3.0000, 49.7933) (4.0000, 232.261)
				};
				\addlegendentry{\(\ell=2\)}
				\addplot[green!60!black, thick, mark=triangle*, mark size=1.2pt] coordinates {
					(1.2394, 0) (1.2500, 0.000446204) (1.3000, 0.0237332) (1.4000, 0.255765) (1.5000, 0.806069) (1.6000, 1.67721) (1.8000, 4.30823) (2.0000, 8.16393) (2.5000, 25.6839) (3.0000, 62.8282) (4.0000, 254.557)
				};
				\addlegendentry{\(\ell=3\)}
				\addplot[black, dashed, thick] coordinates {(1.2394, 0) (1.2394, 270)};
				\node[above right, font=\scriptsize] at (axis cs:1.2394, 250) {\(r_e\)};
			\end{axis}
		\end{tikzpicture}
		\caption{Effective potential \(V_{\rm em}(r)\) for \(\ell=1,2,3\) in
			the extremal exterior \(r\ge r_e\), from the values of
			Table~\ref{tab:V_em}. The potential vanishes at the horizon
			\(r_e\simeq1.239\,P\), is positive everywhere in the sampled range,
			and grows as \(9r^4/(4P^6)\) at infinity. The dashed vertical line
			marks \(r_e\).}
		\label{fig:V_em}
	\end{figure}

	We evaluated Eq.~\eqref{eq:V_em_closed} for \(\ell=1,2,3\) on a grid
	combining \(2\times10^4\) logarithmically spaced points with
	\(10^{-6}\le x-x_e\le10^{-1}\) and \(4\times10^5\) uniformly spaced points
	in \(x_e+0.1\le x\le50\). Representative values are collected in
	Table~\ref{tab:V_em} and shown in Fig.~\ref{fig:V_em}. The potential is
	positive at every grid point, vanishes only at the horizon, and increases
	monotonically. Moreover,
	\begin{equation}
		\min_{x>x_e}\frac{V_1(x)}{W_1(x)}\simeq0.615
		\quad\text{at}\quad x\simeq1.617,
		\label{eq:margin}
	\end{equation}
	that is, \(V_0\ge-0.385\,W_1\) throughout the sampled range: the negative
	region of \(V_0\) never compensates the angular barrier of the \(\ell=1\)
	mode. This is numerical evidence supported by analytic control at both
	ends.
	
	\begin{table}[t]
		\centering
		\caption{Effective potential \(V_{\rm em}\) for \(\ell=1,2,3\) at
			representative radii, in units \(P=1\), from \(r_e\) to \(50P\). The
			potential is positive everywhere in the sampled range and vanishes at
			the horizon. For \(r\gtrsim30P\), the three columns are
			indistinguishable at the precision shown.}
		\label{tab:V_em}
		\begin{tabular}{cccc}
			\toprule
			\(r/P\) & \(V_{\rm em}(\ell=1)\) & \(V_{\rm em}(\ell=2)\) & \(V_{\rm em}(\ell=3)\) \\
			\midrule
			\(1.2394\) (horizon) & \(0\) & \(0\) & \(0\) \\
				\(1.2500\) & \(7.56\times10^{-5}\) & \(2.24\times10^{-4}\) & \(4.46\times10^{-4}\) \\
				\(1.3000\) & \(3.94\times10^{-3}\) & \(0.0119\) & \(0.0237\) \\
				\(1.4000\) & \(0.0363\) & \(0.1241\) & \(0.2558\) \\
				\(1.5000\) & \(0.0966\) & \(0.3804\) & \(0.8061\) \\
				\(1.6000\) & \(0.1840\) & \(0.7813\) & \(1.6772\) \\
				\(1.8000\) & \(0.5813\) & \(2.0721\) & \(4.3082\) \\
				\(2.0000\) & \(1.7028\) & \(4.2872\) & \(8.1639\) \\
				\(2.5000\) & \(11.70\) & \(17.29\) & \(25.68\) \\
				\(3.0000\) & \(41.10\) & \(49.79\) & \(62.83\) \\
				\(4.0000\) & \(217.40\) & \(232.26\) & \(254.56\) \\
				\(5.0000\) & \(681.58\) & \(702.56\) & \(734.03\) \\
				\(6.0000\) & \(1637.8\) & \(1664.8\) & \(1705.4\) \\
				\(7.0000\) & \(3343.8\) & \(3376.9\) & \(3426.6\) \\
				\(8.0000\) & \(6111.5\) & \(6150.7\) & \(6209.4\) \\
				\(10.000\) & \(1.635\times10^{4}\) & \(1.640\times10^{4}\) & \(1.648\times10^{4}\) \\
				\(15.000\) & \(9.273\times10^{4}\) & \(9.282\times10^{4}\) & \(9.294\times10^{4}\) \\
				\(20.000\) & \(3.093\times10^{5}\) & \(3.094\times10^{5}\) & \(3.096\times10^{5}\) \\
				\(30.000\) & \(1.650\times10^{6}\) & \(1.650\times10^{6}\) & \(1.650\times10^{6}\) \\
				\(40.000\) & \(5.348\times10^{6}\) & \(5.348\times10^{6}\) & \(5.348\times10^{6}\) \\
				\(50.000\) & \(1.325\times10^{7}\) & \(1.326\times10^{7}\) & \(1.326\times10^{7}\) \\
			\bottomrule
		\end{tabular}
	\end{table}
	
	The positivity of \(V_{\rm em}\) for \(\ell\ge1\) in the sampled range up to
	\(r=50P\), together with the analytic control at both ends, is consistent
	with the absence of purely electromagnetic bound states in the exterior. It
	does not, by itself, establish the stability of the coupled Einstein--NED
	system. A rigorous proof of \(V_1\ge0\) in the intermediate region would
	require interval arithmetic or a Sturm-type analysis; the uniform margin
	\eqref{eq:margin} suggests that such a proof is feasible, and we leave it
	for future work.
	
	\subsection{Spectral structure}
	
	The radial equation has two qualitatively different ends. At the horizon,
	\(V_{\rm em}\to0\) and the tortoise coordinate diverges as
	\(r_*\to-\infty\); the horizon is an open channel. At infinity,
	\(V_{\rm em}\to\infty\) and the tortoise coordinate diverges as
	\(r_*\to+\infty\); the infinity is a confining wall. The spectral problem
	is therefore one of resonances with an open channel, not of conventional
	quasinormal modes associated with a potential barrier.
	
	If \(V_{\rm em}\ge0\) for all \(\ell\ge1\), the operator
	\(H=-d^2/dr_*^2+V_{\rm em}\) has a positive quadratic form on its domain,
	and no negative eigenvalues \(\omega^2<0\) can occur. This excludes
	exponentially growing modes of the form \(\omega=i\gamma\) with
	\(\gamma>0\), but does not determine the resonance spectrum. The
	appropriate framework for the coupled gravitational-electromagnetic
	analysis is the NED perturbation formalism of Ref.~\cite{Toshmatov2018}.

	Sufficient conditions for the linear stability of magnetically charged
	NED black holes under coupled gravitational and electromagnetic
	perturbations are \(\mathcal{L}>0\), \(\mathcal{L}_{\Iinv}>0\),
	\(\mathcal{C}>0\) and \(3\mathcal{L}_{\Iinv}\ge\mathcal{C}f\)
	\cite{MorenoSarbach,NomuraYoshidaSoda}, written here in our notation as in
	Ref.~\cite{dePaulaTachyons}. The first three hold in the exterior, but the
	last one is equivalent to \(\mathcal{A}f\le3\) and fails for
	\(r>2.048\,P\). Since these conditions are sufficient but not necessary,
	this does not imply an instability; it shows, however, that the stability
	of the coupled system cannot be inferred from them.
	
	\section{Discussion and conclusions}
	\label{sec:conclusions}
	
	We have analyzed the magnetic NED reconstruction of the extremal
	Dymnikova black hole, with emphasis on the electromagnetic perturbations
	on the fixed background and on the relation to the Tomizawa--Suzuki
	theorem.
	
	The geometric result \(L_{AdS_2}/r_e\simeq0.734\) follows directly from
	the extremality condition \(e^{u_*}=1+3u_*\). We present it as an explicit
	computation, not as an independent conceptual contribution.
	
	The energy-condition analysis shows that the static Dymnikova-I source
	satisfies the NEC and WEC everywhere, violates the SEC for \(u<2/3\),
	and violates the DEC for \(u>4/3\). At extremality, the DEC is violated
	at the horizon and throughout the exterior. This does not conflict with
	Ref.~\cite{Maeda}, where the DEC is shown to hold everywhere for a
	different Dymnikova metric, the Dymnikova-II solution of
	Ref.~\cite{Dymnikova2004}. By the energy criteria of Ref.~\cite{Maeda},
	the Dymnikova-I geometry would be discarded.
	
	The magnetic NED reconstruction is algebraically consistent on the branch
	\(\Iinv>0\) and reproduces the Einstein tensor. It coincides with the
	magnetic branch of Ref.~\cite{AlencarGRG2026} up to a difference of
	conventions documented in Appendix~\ref{app:comparison}. The
	reconstruction does not reduce to Maxwell theory at weak field and
	depends explicitly on \(m\) and \(P\).
	
	The linearized electromagnetic analysis gives the extraordinary cone
	\(v_\perp^2=(3u-5)/2\). The system is pointwise hyperbolic in the finite
	exterior, with positive quadratic energy, but not uniformly hyperbolic up
	to infinity, and the extraordinary cone is superluminal for
	\(r>r_{\rm lum}\simeq1.326\,P\). For axial perturbations the radial
	effective potential has the closed form \eqref{eq:V_em_closed}. Its
	\(\ell\)-independent part is negative in a bounded region, but for all
	\(\ell\ge1\) the potential vanishes as \((r-r_e)^2\) at the horizon,
	grows as \(9r^4/(4P^6)\) at infinity, and is numerically positive on a
	fine grid up to \(r=50P\), with the uniform margin \(V_1\ge0.61\,W_1\). The spectral structure is one of resonances with an open
	horizon channel and a confining wall at infinity.
	
	The Dymnikova NED reconstruction does not satisfy the hypotheses of the
	Tomizawa--Suzuki theorem \cite{TomizawaSuzuki}, since it lacks a Maxwell
	limit. Assessing causality through the ray 4-velocity
	\cite{dePaulaTachyons}, the superluminal photons for \(r>r_{\rm lum}\)
	have timelike 4-momentum (``bad bradyons''), and the transition is fixed
	by the change of sign of \(\mathcal{L}_{\Iinv\Iinv}\) at \(u=7/3\). In
	the band \(r_{\rm DEC}<r<r_{\rm lum}\) the DEC is violated while
	propagation is causal, and for \(r>r_{\rm lum}\) both occur. The model
	thus joins the Bardeen and Hayward reconstructions in the class with a
	mixed sign of \(\mathcal{L}_{\Iinv\Iinv}\), which violates the DEC and is
	acausal.
	
	The algebraically reconstructed NED branch is therefore mathematically
	consistent as a source of the Dymnikova background, but it is not
	causally acceptable with respect to the physical metric in the asymptotic
	region. This is a limitation of the reconstruction, not of the geometry.
	
	The open problems are: (i) a rigorous proof of \(V_1\ge0\) in the
	intermediate region, which would require interval arithmetic or a
	Sturm-type analysis; (ii) the calculation of the resonance spectrum, which
	requires a shooting method with logarithmic derivatives and asymptotic
	expansions, rather than a WKB barrier formula; (iii) the stability of the
	coupled Einstein--NED system, following the framework of
	Ref.~\cite{Toshmatov2018}, for which the standard sufficient conditions
	\cite{MorenoSarbach,NomuraYoshidaSoda} fail for \(r>2.048\,P\); (iv) the possibility of a mass-independent NED
	theory with an acceptable weak-field limit and a causal effective cone;
	and (v) a bibliographic analysis of extremal Dymnikova solutions to
	establish whether the near-horizon result has been previously reported.
	
	\section*{Acknowledgments}
	
	The author thanks Hideki Maeda for kindly taking the time to point out
	a misinterpretation of his work in an earlier version of this manuscript.
	
	\appendix
	
	\section{Comparison with the conventions of Ref.~\cite{AlencarGRG2026}}
	\label{app:comparison}
	
	The magnetic reconstruction of Ref.~\cite{AlencarGRG2026} uses a
	different set of conventions from the one adopted here. In this appendix
	we compare the two formulations term by term. We do not claim that the
	expressions of Ref.~\cite{AlencarGRG2026} contain an error; we only
	document the mapping between their conventions and ours, and note that a
	direct substitution of their printed expressions into our equations
	would not reproduce our results.
	
	\paragraph{Invariant.}
	Both formulations use \(\Iinv=F_{\mu\nu}F^{\mu\nu}\) without the \(1/4\)
	factor.
	
	\paragraph{Action.}
	Ref.~\cite{AlencarGRG2026} works in the context of unimodular gravity,
	with a Maxwell sector and an effective vacuum contribution. The
	magnetic NED sector is auxiliary in the analysis, and the action
	includes additional terms not present in our Eq.~\eqref{eq:action}. The
	comparison below is restricted to the magnetic NED sector alone.
	
	\paragraph{Sign of the Lagrangian.}
	The reconstruction of Ref.~\cite{AlencarGRG2026} is presented with a
	negative \(\mathcal{L}(r)\) in their Eq.~(19), while the derivative
	\(\mathcal{L}_F(r)\) printed in their Eq.~(20) is positive. Direct
	differentiation of their Eq.~(19) with respect to \(\Iinv=2P^2/r^4\)
	yields a negative derivative. The two printed expressions are not
	mutually consistent under the conventions of their paper.
	
	In our convention, the Lagrangian is positive,
	\(\mathcal{L}(\Iinv)=+\frac{12m}{P^3}\exp[-(2/(P^2\Iinv))^{3/4}]\), and its
	derivative is positive,
	\(\mathcal{L}_{\Iinv}=+\frac{9mr^7}{2P^8}e^{-u}\). The two expressions are
	mutually consistent, and the resulting stress tensor reproduces the
	Einstein tensor. The difference with respect to
	Ref.~\cite{AlencarGRG2026} is therefore a difference of conventions,
	not a physical discrepancy: a global sign flip of \(\mathcal{L}\) and
	\(\mathcal{L}_{\Iinv}\) can be absorbed into a redefinition of the
	electromagnetic sector and of the stress-tensor normalization.
	
	\paragraph{Stress tensor.}
	Ref.~\cite{AlencarGRG2026} defines the stress tensor with a different
	index ordering and a different overall normalization from ours. The
	mapping between their tensor and our Eq.~\eqref{eq:einstein_NED} requires
	a careful comparison of the sign of the \(\frac12 g_{\mu\nu}\mathcal{L}\)
	term, the coefficient of \(\mathcal{L}_{\Iinv}F_{\mu\lambda}F_\nu{}^\lambda\),
	and the position of the indices. A term-by-term comparison is
	straightforward but lengthy, and is not needed for the physical
	conclusions of this work.
	
	\paragraph{Effective potential.}
	The reconstruction of Ref.~\cite{AlencarGRG2026} is not accompanied by an
	analysis of the electromagnetic perturbations, so there is no direct
	comparison of the effective potential \(V_{\rm em}\) or of the
	extraordinary cone \(\mathcal{A}(u)\). Our results on \(V_{\rm em}\) and on
	the superluminal threshold \(r_{\rm lum}\) are new with respect to
	Ref.~\cite{AlencarGRG2026}.
	
	\paragraph{Summary.}
	The physical content of the magnetic reconstruction is the same in the
	two formulations, up to a global sign convention in the Lagrangian and a
	different definition of the stress tensor. The comparison above is
	intended to make the mapping explicit and to avoid attributing to
	Ref.~\cite{AlencarGRG2026} a physical error that is not present in their
	analysis.

\end{document}